\documentstyle[aps,preprint,tighten,epsf]{revtex}

\begin{document}

\title{\vspace*{1in}
       SEARCHING FOR THE MSW ENHANCEMENT}

\author{James M. Gelb, Waikwok Kwong, and S. P. Rosen}

\address{Department of Physics, The University of Texas at Arlington, 
         Arlington, Texas 76019}

\date{\today}

\maketitle

\begin{abstract}
We point out that the length scale  associated with the MSW effect is the
radius of the Earth. Therefore to verify matter enhancement of neutrino
oscillations, it will be necessary to study neutrinos passing through the
Earth. For the parameters of MSW solutions to the solar neutrino problem, the
only detectable effects occur in a narrow band of energies from 5 to 10 MeV.
We propose that serious consideration be given to mounting an experiment at a
location within 9.5 degrees of the equator. 
\end{abstract}


\narrowtext

Because the small angle MSW solution~\cite{HL,RG,BK,TU} gives the
best fit to existing solar neutrino data~\cite{SOLDAT}, it could be the
ultimate solution to the solar neutrino problem\cite{JNB}, and independent
means of verifying it become important. The scale of  the MSW
effect~\cite{MSW} is set by the radius of the Earth (see below),  and so the
{\it day-night} effect~\cite{CBW} is probably the only means of demonstrating
unambiguously that the passage of neutrinos through matter can enhance
neutrino oscillations with small {\it in vacuo} mixing angles. It is
unfortunate that the total rate for the Kamiokande experiment~\cite{KAMDN}
shows no clear evidence for a nighttime signal greater than the daytime one,
and that the preferred fit to the total rates of  all four solar neutrino
experiments lies outside the region of parameter space where such an effect is
expected~\cite{HL,CBW}.  Here we want to argue that the best place to look for
the {\it day-night} effect is in a band of neutrino energies from about 5 to
10 MeV and at a new location, close to the Equator. 

When we rewrite the famous MSW enhancement condition~\cite{MSW}
 $\sqrt{2} G_F N_e =(\Delta m^2/2E) \cos 2\theta$,
so that $E/\Delta m^2$ is expressed as a length in meters, we find that the
controlling scale happens to be the radius of the Earth $R_E$:
  \begin{eqnarray}
  \frac{E}{\Delta m^2} = \frac{6.7\times 10^6}{\rho_E (Z/A)}\cos 2\theta=
\frac{1.05R_E}{\rho_E (Z/A)} \cos 2\theta.
  \end{eqnarray}
The density of  the Earth, $\rho_E$, varies from about 3 gm/cc at the surface
to 13 gm/cc in the core~\cite{DEN}; and the ratio ($Z/A$) is of order 0.5
throughout. Therefore, in order for a neutrino oscillation to be enhanced in
passage through the Earth, its {\it in vacuo} oscillation length must be at
least a significant fraction of $R_E$. 

To gain some insight into possible matter oscillations, it is instructive to
apply the parameters of typical MSW fits to a  model of the Earth. The small
angle MSW solution is typified by~\cite{HL,BK}: 
  \begin{eqnarray}
  \Delta m^2 = 6 \times 10^{-6} \rm~eV^2, \qquad
  \sin^2 2 \theta = 0.007.
  \end{eqnarray}
The density  $\rho_E$~\cite{DEN}, as a function of the distance $xR_E, (0 \leq
x \leq 1)$ from the center of the Earth, decreases monotonically from 13 gm/cc
at $x = 0$ to 10 gm/cc at $x = 0.55$ in the core region, and after a sharp
discontinuity, decreases monotonically  again in the mantle from 6 gm/cc at $x
= 0.55$ to 3 gm/cc at $x = 1$. 
 
Neutrinos traveling through the Earth will satisfy the enhancement condition
in the core when their energies fall within the narrow band~\cite{BW1} of
$6.5 {\rm~MeV} \leq E \leq 8{\rm~MeV}$, and in the mantle when they fall into
a wider, higher energy band of $13.5 {\rm~MeV} \leq E \leq 27 \rm~MeV$.

The size of the enhancement region is~\cite{RG} $\Delta R = h_0 \tan 2\theta$,
where $h_0$ is the reciprocal of the logarithmic derivative of the density;
for the core, it is 
  \begin{eqnarray}
  \Delta R_{\rm core} \approx 2 R_E \tan 2\theta \, \, \approx 0.17 R_E,
  \end{eqnarray}
and for the mantle
  \begin{eqnarray}
  \Delta R_{\rm mantle} \approx 0.75 R_E \tan 2\theta \,\, \approx 0.06 R_E.
  \end{eqnarray} 
Both regions are considerably smaller than the corresponding oscillation
lengths at the points of enhancement, $L_m \approx (5.3\cot 2\theta/
\rho_0)R_E$, and so the enhancements are nonadiabatic in
character~\cite{RG,TU}. Since the energy band $6.5 {\rm~MeV} \leq E \leq
8{\rm~MeV}$ overlaps the $^8$B solar neutrino spectrum, we shall concentrate
our attention on it from now on. 

In passing, we note that the large angle MSW solution, typified by the
oscillation parameters~\cite{HL,BK} 
  \begin{eqnarray}
  \Delta m^2  = 3 \times10^{-5} \rm~eV^2, \,\,\,\, \sin^2 2\theta  =  0.7, 
  \end{eqnarray}
predicts enhancement at energies well outside the solar spectrum (except
perhaps for the hep branch). They are, in units of MeV: $18.5\leq E \leq 24$,
and $40\leq E \leq 80$, for the core and mantle of the Earth respectively. 

For the enhancement mechanism to have a significant effect on the $\nu_e$
survival probability of neutrinos in the $6.5 {\rm~MeV} \leq E \leq
8{\rm~MeV}$ energy band, the neutrinos must travel a distance $\Delta R_c$ or
more through the vicinity of the enhancement density. The discontinuity in
density between the core and mantle occurs at $x = 0.55$ and subtends a
half-angle of 33$^\circ$ at the surface of the Earth. Therefore paths must be
within 33$^\circ$ of the vertical to produce significant effects. To
illustrate the point, imagine these neutrinos traveling through the Earth on
a path which makes an angle $\phi$ with the vertical at the site of the
detector. When $\phi$ is less than 29.88$^\circ$, the path length in the core
will be greater than $\Delta R_{\rm core} \approx 0.17 R_E$ as given above, 
and there will be a significant effect on oscillation probabilities. It falls
off dramatically as $\phi$ reaches the 30$^\circ$ mark. 

Another subtle effect is that the neutrino energy corresponding to the largest
change in oscillation probability will increase with $\phi$. Enhancement
energies (see Eqs.~1 and 2) vary inversely with density, and the maximum
density along any path decreases as
$\phi$ increases. Since interaction cross sections increase with energy, this
effect may be important in choosing a path to maximize neutrino interaction
rates. 

To study the magnitude of these effects, we have performed a series of
numerical calculations which follow  $^8$B solar neutrinos from the point of
origin in the Sun through different chords of the Earth. The $\nu_e$ survival
probability is very sensitive to the relative phase between the neutrino mass
eigenstate components when they arrive at Earth. Since the neutrinos are
produced over a region of the solar core which is about 10 $R_E$, and since
their {\it in vacuo} oscillation length is only a fraction of $R_E$, we
average over the relative phase to calculate the $\nu_e$ survival
probability~\cite{MS}. 

Our results are shown in Fig.~1. There is a significant increase in survival
probability for neutrinos between 5 and 10 MeV for all angles less than
33$^\circ$. The probability increases by almost a factor of 2, and the energy
at which the largest increase occurs tends to move up with $\phi$. By about 10
MeV there is no longer a difference between day and night. A slight increase
in the neighborhood of 12 MeV corresponds to a small enhancement in the mantle
of the Earth. 

To carry out the search for Earth enhancement, it will be best to detect
neutrinos through their charged-current interactions with as heavy a target as
possible: the heavier the target, the more closely the recoil electron energy
follows the energy of the incident neutrino. In neutrino--electron scattering,
for example, the recoil electron energies are almost equally distributed
amongst the values kinematically allowed by  the incident neutrino energy, and
they are therefore spread over a considerable range~\cite{GBR}. In
neutrino--deuteron charged-current interactions, the electron energy is sharply
peaked at its maximum value, but does have a tail extending to lower
energies~\cite{KR}. This implies that Earth enhancement can noticeably change
the {\it shape} of the recoil spectrum in neutrino--deuteron interactions, but
not in $\nu$--$e$ scattering. Experiments like SNO~\cite{SNO} may therefore be
better suited to this measurement than ones like Super Kamiokande~\cite{SK},
although the latter has great statistical power. Heavier targets such as
$^{12}$C and $^{16}$O may be even more suitable, especially if the recoil
electron can be detected and its energy measured in real time. 

We can demonstrate these points by calculating the differential cross sections
for neutrino interactions as a function of the angle $\phi$ through the Earth
and comparing them with daytime cross sections (see Fig.~2). At $\phi
=0^\circ$, the deuterium differential cross section increases significantly
for recoil electrons with kinetic energies between 4 and 6 MeV,  and at $\phi
=20^\circ$,  the increase shifts to higher energies between 5.5 and 7.5 MeV.
By contrast, the  elastic scattering cross section shows a small, essentially 
uniform increase for all recoil energies below 7 MeV, and the increase does
not vary much with $\phi$. 

We can compare nighttime rates, $N$, as a function of $\phi$ and express the
difference, $N - D$, from the daytime rate, $D$, as a percentage of $D$. To
take shape effects into account, we use three different energy ranges for the
recoil kinetic energy of the electron: (i) 0--$T_{max}$, the maximum recoil
kinetic energy; (ii) 5 MeV--$T_{max}$; and (iii) 5--10 MeV. Our results for
deuterium interactions and elastic scattering are shown in Fig.~3. For elastic
scattering, there is roughly a $15\%$ increase for angles up to 27$^\circ$ and,
as expected, this behavior is much the same for all three energy ranges. For
deuterium, however, the changing shape of the recoil spectrum with angle $\phi$
shows up when we cut out recoil energies below 5 MeV, and the behavior in
ranges (ii) and (iii) is different from that in (i). 

Location is an important consideration maximizing enhancement effects. There
are two aspects: the {\it coverage time} in any one year that solar neutrinos
must pass through the core of the Earth to reach the detector; and the minimum
angle $\phi_{min}$ of the chords they travel. To study these questions we need
to calculate the hours during which rays from the Sun will pass through the
core of the Earth ($\phi \leq 33^\circ$) on any day of the year for a latitude
$\ell$. Let $\Omega$ be the orbital phase of the Earth around the Sun ($\Omega
= 0^\circ$ during the Summer Solstice, 90$^\circ$ for the Autumnal Equinox, and
so on). The declination of the Sun is $\delta = 23.5^\circ \cos\Omega$, and
hence the Sun is located at an angle of $(90^\circ - \ell + \delta)$ from the
southern horizon at high noon. If  we define the hour angle $h$ to run from
($-12$) to (+12), with $h = 0$ as high noon, we can determine the critical hour
$h_c$ when the Sun passes through the critical angle $\phi_c = 33^\circ$ and
the duration $D_h$, in hours, that the Sun is blocked by the core: 
  \begin{eqnarray} 
  h_c & = &\frac{24}{2\pi}\cos^{-1}\left(-\frac{\cos \theta_c + \sin\delta  
  \sin\ell}{\cos \delta \cos\ell}\right); \nonumber\\
  D_h &=& 2\,(12-h_c). 
  \end{eqnarray}
It is simple to show that the lowest latitude for which the rays from the Sun
will never pass  through the core of the Earth is $\ell_n =\phi_c + 
\delta_{max}= 33^\circ + 23.5^\circ = 56.5^\circ$; it can also be shown that
the highest latitude for which solar neutrinos will pass through the core for
some time during each day of the year is $\ell_h = \phi_c - \delta_{max} =
9.5^\circ$. 

Figure 4 shows the coverage time per day as a function of the seasons of the
year. At 0$^\circ$ latitude, the coverage time oscillates between a maximum of
almost 4.5 hours during the Vernal and Autumnal Equinoxes, and a minimum of
about 3.25 hours at the Summer and Winter Solstices; altogether the coverage
time amounts to 14.3\% of the year. As the latitude increases, this percentage
decreases and, once $\ell \geq 9.5^\circ$, the part of the year for which there
is no coverage grows steadily. By 25$^\circ$, coverage occurs only in the Fall
and Winter and at the Winter Solstice, the daily coverage time is almost 5
hours. This has the interesting effect of doubling the coverage during the
relevant part of the year. For example, at 30$^\circ$, the total coverage time
is $7.5\%$ of the full year, but 15\% of the Fall and Winter; thus comparing
night and day during the Fall and Winter alone enhances the overall signal. 

Higher latitudes also cover a smaller range of angles $\phi$. At
12 midnight the angle $\phi$ at latitude $\ell$ is ($\delta + \ell$), its
smallest value for the night. For a given latitude less than $23.5^\circ$,
the angle ($\delta + \ell$) will vanish at some time in the winter and $\phi$
will run the full range from $0^\circ$ to $33^\circ$. For $\ell \geq
23.5^\circ$,  the minimum angle $\phi_{min}$ never vanishes and increases with
latitude. By $45^\circ$, we are running out of the most effective enhancement
region of the core (see Fig.~3). 

These factors of coverage time and angular range make it clear that latitudes
close to the Equator are the most favorable for maximal day-night effects. 
Present experiments~\cite{SOLDAT} are all at latitudes greater than 30$^\circ$
and coverage is restricted to the Fall and Winter: Super Kamiokande, at a
latitude of 36.4$^\circ$, is covered by the core $7\%$ of the year and has 
$\phi_{min}$ of 13$^\circ$; SNO is at 46.5$^\circ$ and is covered 3.5\% with
$\phi_{min}$ of 23$^\circ$; and Borexino, at about 42$^\circ$, is covered 5\%
of the year with $\phi_{min}$ of 18.5$^\circ$. In all three cases, coverage is
restricted to the Fall and Winter. We therefore propose that a search be made
for a new site, preferably within 9.5$^\circ$ of the Equatorial region with
sufficient overburden, 3000 mwe or more, and infrastructure to mount a major
experiment. 

To illustrate the experimental requirements necessary for the observation of a
significant  effect, we consider a charged-current deuterium experiment located
at the Equator. During the $14.3\%$ of the coverage time in any one year
(``nighttime"), the event rate will be greater by roughly $20\%$ than that
during the corresponding $14.3\%$ that the detector is diametrically opposite
the covered locations (``daytime"). Therefore, if $X$ events are seen in the
``daytime", then 1.2$X$ will be seen in the ``nighttime". If we assume that
systematic errors are equal to statistical ones and ask for a 3$\sigma$ effect,
then $X$ = 900. In other words we require about 6300 events for the full year,
or  17 events per (24 hour) day. This is roughly twice the SNO rate and 40
times the rate of the Davis $^{37}$Cl experiment~\cite{JNB}. Thus we would
require 2 kilotons of heavy water, or 24 kilotons of cleaning fluid (assuming
that $^{37}$Cl events can be seen in real time). A 5$\sigma$ effect would
require almost three times these quantities, or three times the duration, or
some combination of increased size and duration. 

In conclusion, we have outlined the case and the scope for a new solar neutrino
experiment dedicated to determining whether the matter enhancement of neutrino
oscillations does indeed occur. We believe that this experiment is a necessary
sequel to the experiments presently operating, or coming online . If the small
angle MSW should continue to give the best fit to existing and forthcoming
solar neutrino data, the observation of a night-day effect~\cite{CBW} will be
the conclusive evidence that it is indeed the true solution to the solar
neutrino problem. 

This work is supported in part by the U. S. Department of Energy Grant
No.~DE-FG05-92ER40691. One of the authors (SPR) is grateful to Dr.~Geoffrey
West for a valuable remark in the early stages of this work, and to Los Alamos
National Laboratory for its hospitality at the 1995 Santa Fe Workshop {\it
Massive Neutrinos and Their Implications}, July 24--August 11, 1995.

\begin{figure}
\vspace*{1in}
\epsfxsize=\hsize
\epsfbox{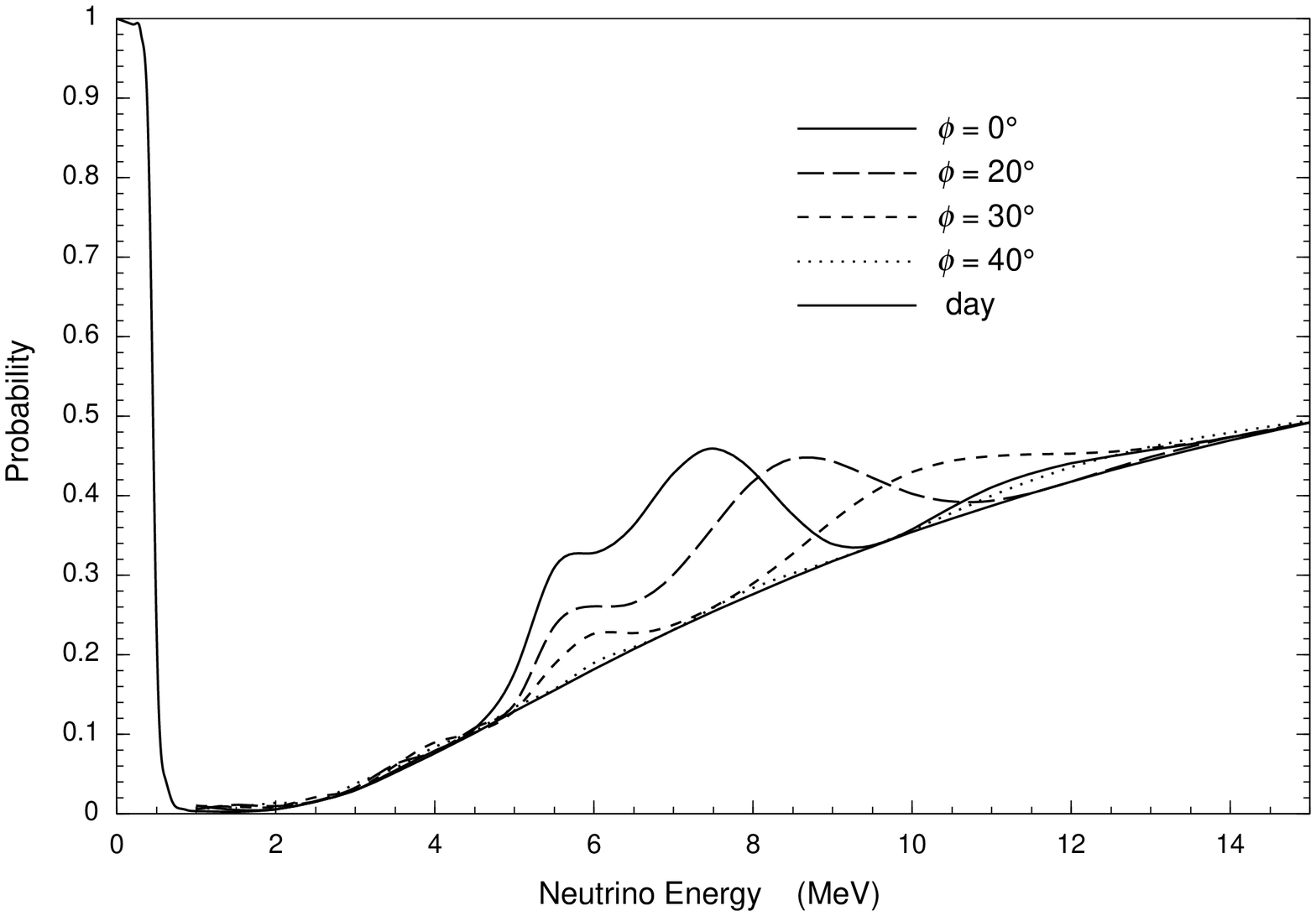}
\caption{Survival probability $P(\nu_e{\to}\nu_e)$ for $^8$B neutrinos before
(day) and after passing through the Earth at various angles (night).} 
\end{figure}

\begin{figure}
\epsfxsize=0.9\hsize
$$\epsfbox{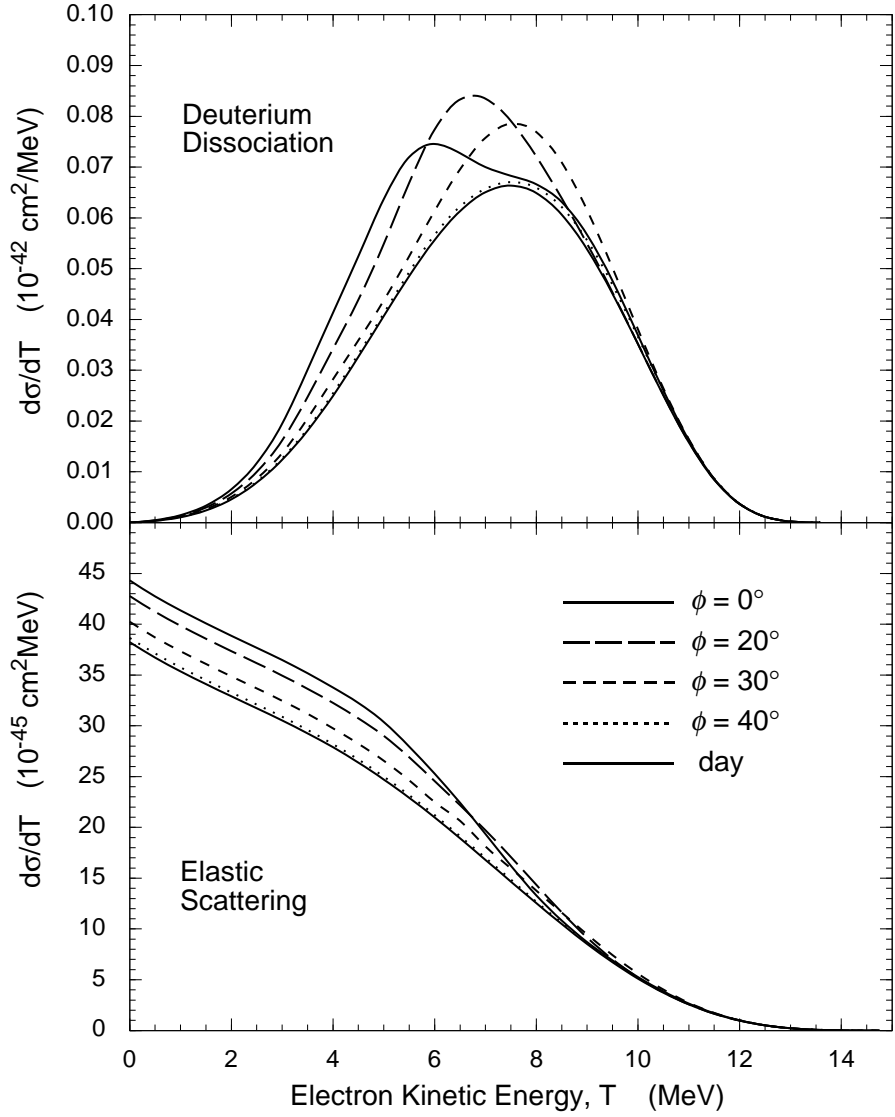} $$
\caption{Differential cross sections averaged over the $^8$B spectrum, before
and after passing through the Earth.} 
\end{figure}

\begin{figure}
\epsfxsize=0.9\hsize
$$\epsfbox{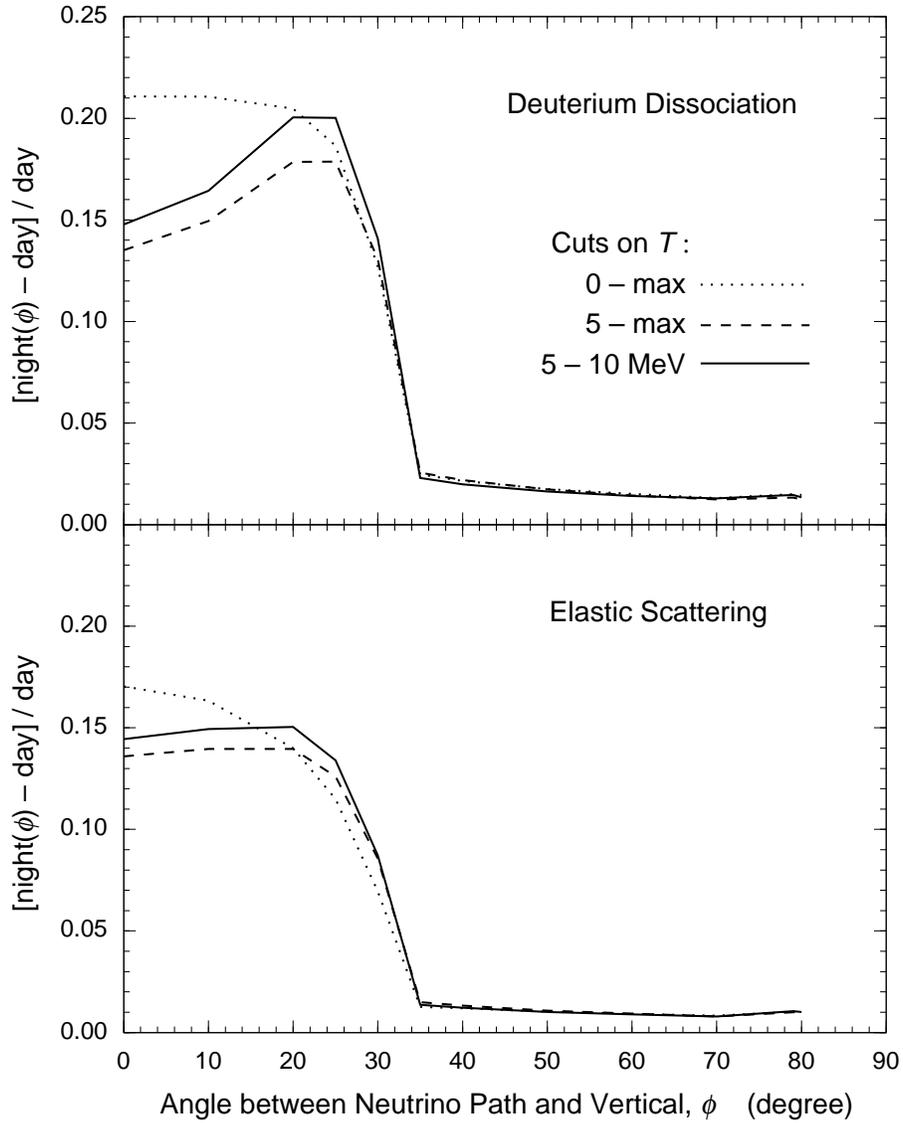} $$
\caption{Fractional excess of event rates at night as a function of the
neutrino path.} 
\end{figure}

\begin{figure}
\epsfxsize= 0.9\hsize
$$\epsfbox{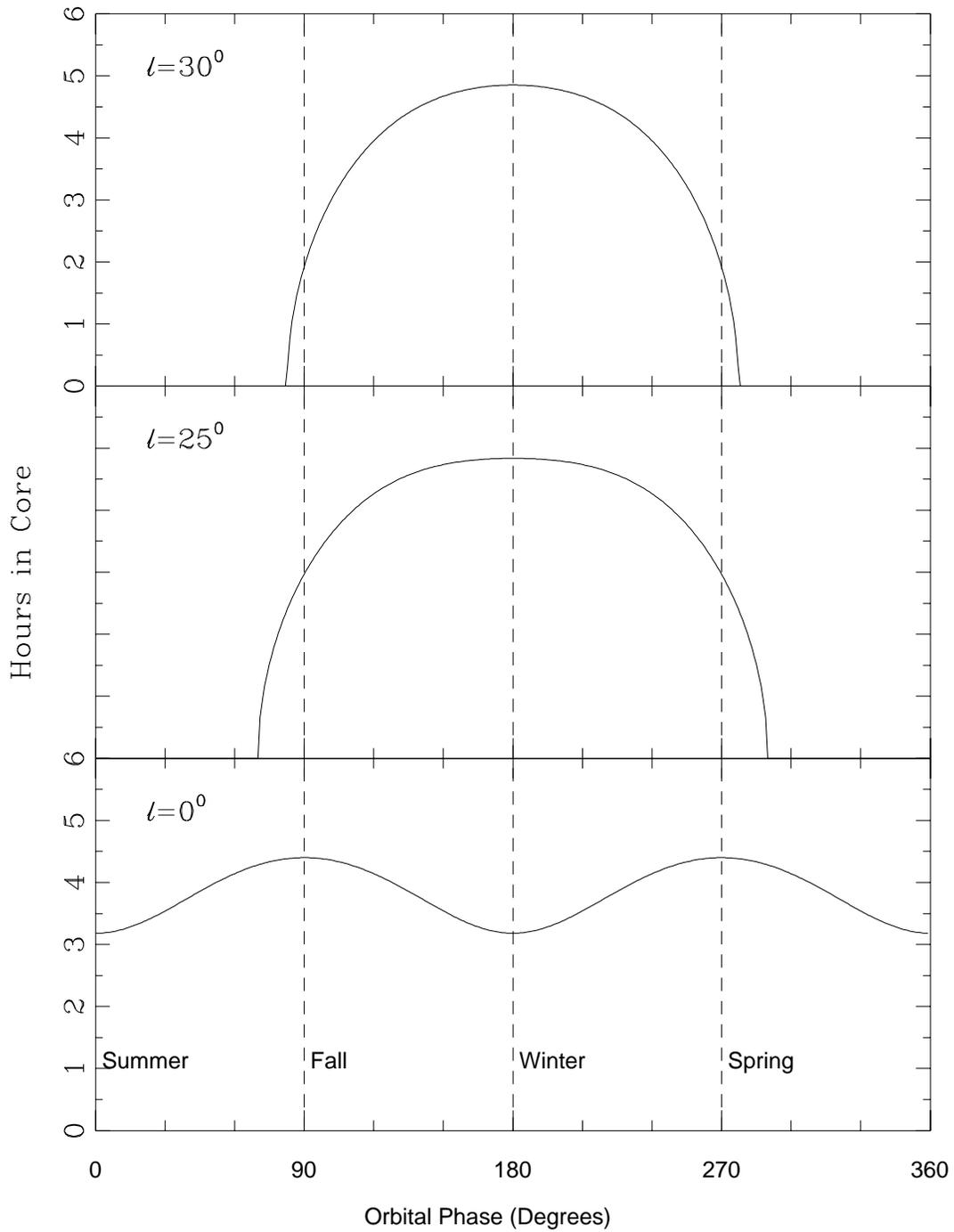} $$
\caption{Number of hours per night that solar neutrinos pass through the
Earth's core: annual variations for various latitudes.} 
\end{figure}

\end{document}